\documentclass[preprint]{aastex}
\newcommand{\ea}{et al.}

\newcommand{\bfi}{\begin{figure}[htb]} 
\newcommand{\bpfi}{\begin{figure}[p]}

\def\hii{\relax \ifmmode {\rm H\,{\sc ii}}\else H\,{\sc ii}\fi}

\def\fdg{\hbox{$.\!\!^\circ$}}

\def\farcm{\hbox{$.\mkern-4mu^\prime$}}
\def\farcs{\hbox{$.\!\!^{\prime\prime}$}}
\def\degd#1.#2{ #1\fdg#2 }                 % degrees over decimal point
\def\mind#1.#2{ #1\farcm#2 }               % minutes over decimal point  
\def\secd#1.#2{ #1\farcs#2 }               % seconds over decimal point 

\def\aj{AJ}                   % Astronomical Journal
\def\apj{ApJ}                 % Astrophysical Journal
\def\apjl{ApJ}                % Astrophysical Journal, Letters
\def\apjs{ApJS}               % Astrophysical Journal, Supplement
\def\ao{Appl.~Opt.}           % Applied Optics
\def\aap{A\&A}                % Astronomy and Astrophysics
\def\mnras{MNRAS}             % Monthly Notices of the RAS
\def\physrep{Phys.~Rep.}   % Physics Reports
\usepackage{graphicx}
\usepackage{color}

\def\today{\ifcase\month\or
 January\or February\or March\or April\or May\or June\or
 July\or August\or September\or October\or November\or
 December\fi\space\number\day, \number\year}

\def\todmy{\number\day\space\ifcase\month\or
 January\or February\or March\or April\or May\or June\or
 July\or August\or September\or October\or November\or
 December\fi\space\number\year}

\shorttitle{Ansae in barred galaxies}
\shortauthors{Martinez-Valpuesta et al.}

\begin {document}
%% LaTeX will automatically break titles if they run longer than
%% one line. However, you may use \\ to force a line break if
%% you desire.

\title{A Morphological and Statistical Analysis of Ansae in Barred Galaxies}

%% Use \author, \affil, and the \and command to format
%% author and affiliation information.
\author{I.~Martinez-Valpuesta\footnote{e-mail address: imv@iac.es}}
\affil{Instituto de Astrof\'\i sica de Canarias, E-38200 La Laguna,
  Tenerife, Spain \\
LAM, Observatoire Astronomique de Marseille Provence, 2 Place Le
  Verrier, F-13004 Marseille, France}

\author{J.~H.~Knapen}
\affil{Instituto de Astrof\'\i sica de Canarias, E-38200 La Laguna,
  Tenerife, Spain }
  \and
\author{R.~Buta}
\affil{Department of Physics and Astronomy, Box 870324, University of Alabama, Tuscaloosa,
AL 25487, USA}

%% Mark off your abstract in the ``abstract'' environment. In the manuscript
%% style, abstract will output a Received/Accepted line after the
%% title and affiliation information. No date will appear since the author
%% does not have this information. The dates will be filled in by the
%% editorial office after submission.

\begin{abstract}

Many barred galaxies show a set of symmetric enhancements at the ends
of the stellar bar, called {\it ansae}, or the ``handles'' of the
bar. The ansa bars have been in the literature for some decades,
but their origin has still not been specifically addressed, although, they could be related to the growth process of bars.  But even though ansae have been known for a long time, no statistical analysis of their relative frequency of 
occurrence has been performed yet. Similarly, there has been no
study of the varieties in morphology of ansae even though
significant morphological variations are known to characterise the
features. In this
paper, we make a quantitative analysis of the occurrence of ansae in
barred galaxies, making use of {\it The de Vaucouleurs Atlas of
Galaxies} by Buta and coworkers. We find that $\sim 40\%$ of 
SB0's show ansae in their bars, thus confirming that ansae are
common features in barred lenticulars. The ansa frequency decreases
dramatically with later types, and hardly any ansae are found in
galaxies of type Sb or later. The bars in galaxies with ansae are
stronger in the median than those in galaxies without ansae, but the
presence of inner and outer rings is not related to the presence of
ansae. Implications of these results and theories for the possible
origin of ansae are discussed briefly.

\end{abstract}


\section{Introduction}

Bars in disk galaxies are both common and long-lived. Studies based on
optical images have established that about $1/3$ of all disk galaxies
are strongly barred, and an additional $1/3$ are moderately barred
(e.g., Sandage 1961, Sellwood \& Wilkinson 1993). When using near-infrared (NIR)
imaging, which is more sensitive to the older stellar populations that
predominate in bars, the fraction of barred galaxies rises to close to
$80\%$ of local disk galaxies (Knapen, Shlosman \& Peletier 2000;
Eskridge et al. 2000; Grosb\o l et al. 2002; Marinova \& Jogee 2006; 
Men\'endez-Delmestre \ea 2007). From studies of
images obtained with the {\it Hubble Space Telescope}, 
it has recently been determined that the fraction of strong bars remains
essentially unchanged from intermediate redshifts, of $z\sim 1.2$, to
the present day (e.g., Sheth et al. 2003; Jogee et al. 2004; 
Elmegreen et al. 2004; Zheng \ea 2005). This agrees with numerical simulations, 
which show that bars are indeed long-lived phenomena
(e.g., Debattista \& Sellwood 2000; Athanassoula \& Misiriotis 2002; 
Valenzuela \& Klypin 2003; Shen \& Sellwood 2004; Debattista et al. 2004; 
Martinez-Valpuesta, Shlosman \& Heller 2006).

An early approach to the shapes of stellar bars considered them, in
principle, to be ellipsoidal features in disk galaxies. Consequently,
the isodensity contours of bars have traditionally been fitted with
ellipses. Parameters to account for the boxiness of observed stellar
bars were later introduced by Nieto \& Bender (1990), who considered
the elliptical fits unsatisfactory. The departure of bar shapes from
strictly elliptical has also been noticed in numerical simulations
(e.g., Sparke \& Sellwood 1987). An improved method, based on
generalised ellipses, has been tested successfully and applied to
both numerical and observed bars (Athanassoula et al. 1990).

The existence of an important diversity in the shapes and structure of observed and
simulated bars is widely acknowledged today (Ohta, Hamabe \& Wakamatsu 1990; 
Regan \& Elmegreen 1997; Aguerri, Beckman \& Prieto 1998, 2000; Odewahn et al. 2002). 
The deviations from pure
elliptical shapes vary from mild (and generally weak) oval distortions
dominated mainly by the $m$=2 Fourier term to highly elongated and
rectangular bars (which are generally stronger) showing significant
higher order terms. Not all the
morphological features in observed barred galaxies are, however, easy
to identify and classify. We are specifically interested here in the
outer regions of stellar bars, which give rise to their pronounced
shapes. This is the region where stellar bars in spirals connect to the spiral
arms. Physically, this region could lie between the inner 4:1 ultraharmonic
resonance (UHR) and the corotation resonance (CR) radius of the bar (e.g., 
Aguerri, Debattista \& Corsini 2003, and references therein) . Various
families of exotic orbits are found here, which are frequently
unstable, and so their actual particle population is
unknown. Different features, such as the ansae we concentrate on here,
frequently appear in this region.

\begin{figure}
\begin{center}
\includegraphics[scale=0.5]{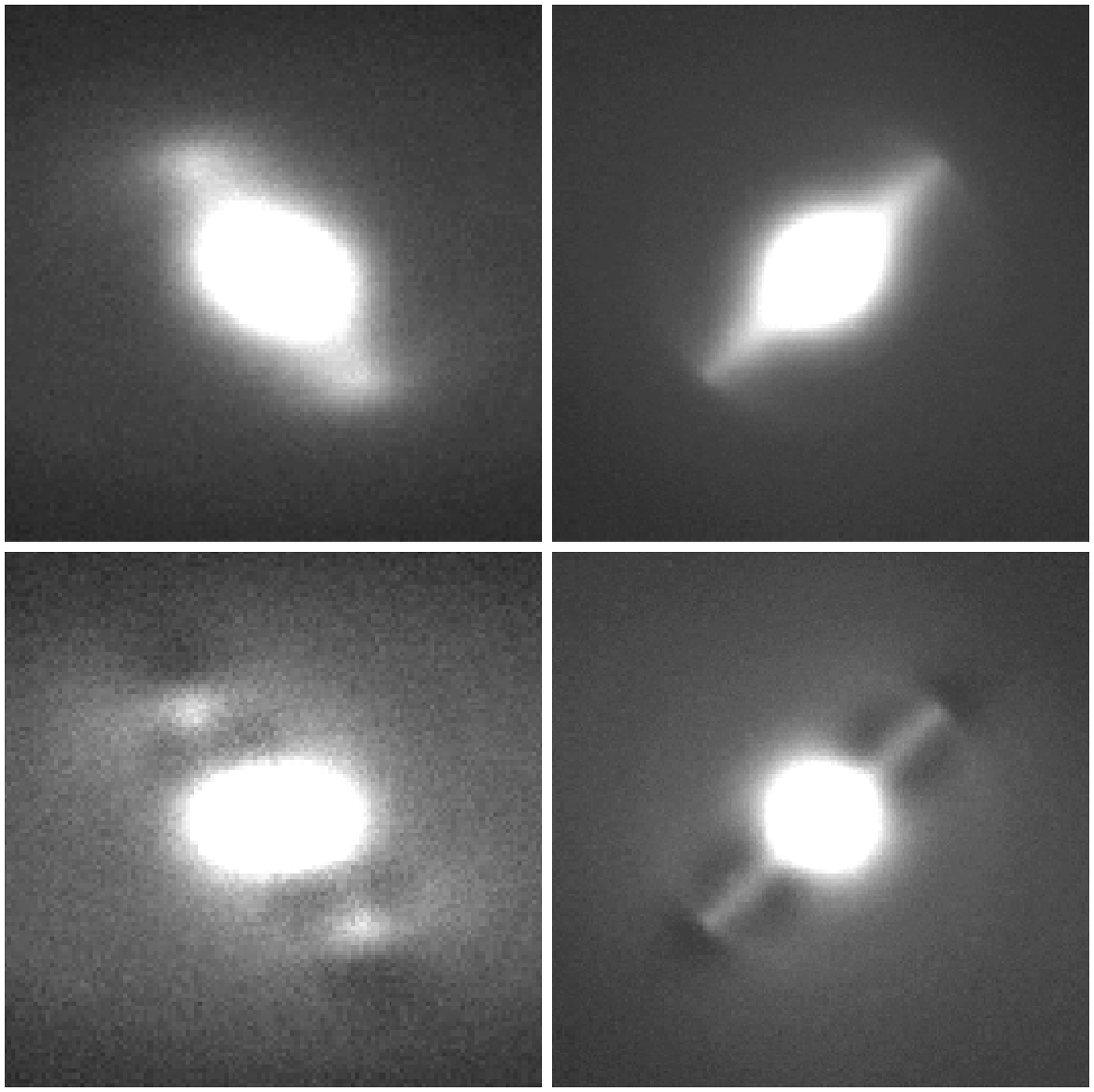}
\caption{Illustration of two strongly-barred early-type galaxies, one
having ansae (NGC 2983, left panels) and the other without ansae
(NGC 4643, right panels). The upper left panel shows a $V$-band
image of NGC 2983 obtained with the CTIO 1.5m telescope by
Buta \& Crocker (1991). The upper right panel
shows an $i$-band image of NGC 4643 from the Sloan Digital Sky
Survey. The lower panels show the same images after a smooth Ferrers
bar model has been subtracted. These models were
derived from two-dimensional bulge/disk/bar decompositions
using the program and methods described by
Laurikainen, Salo \& Buta (2005).
The bar-subtracted images reveal very clearly
that NGC 2983 has density enhancements at the ends of its bar
while NGC 4643 shows no trace of such features.
This difference is not due to the bar model used, which is in
any case only an imperfect representation of both bars.
}
\label{fig:NGC4643}
\end{center}
\end{figure}

\begin{figure}
\begin{center}
\includegraphics[scale=0.5]{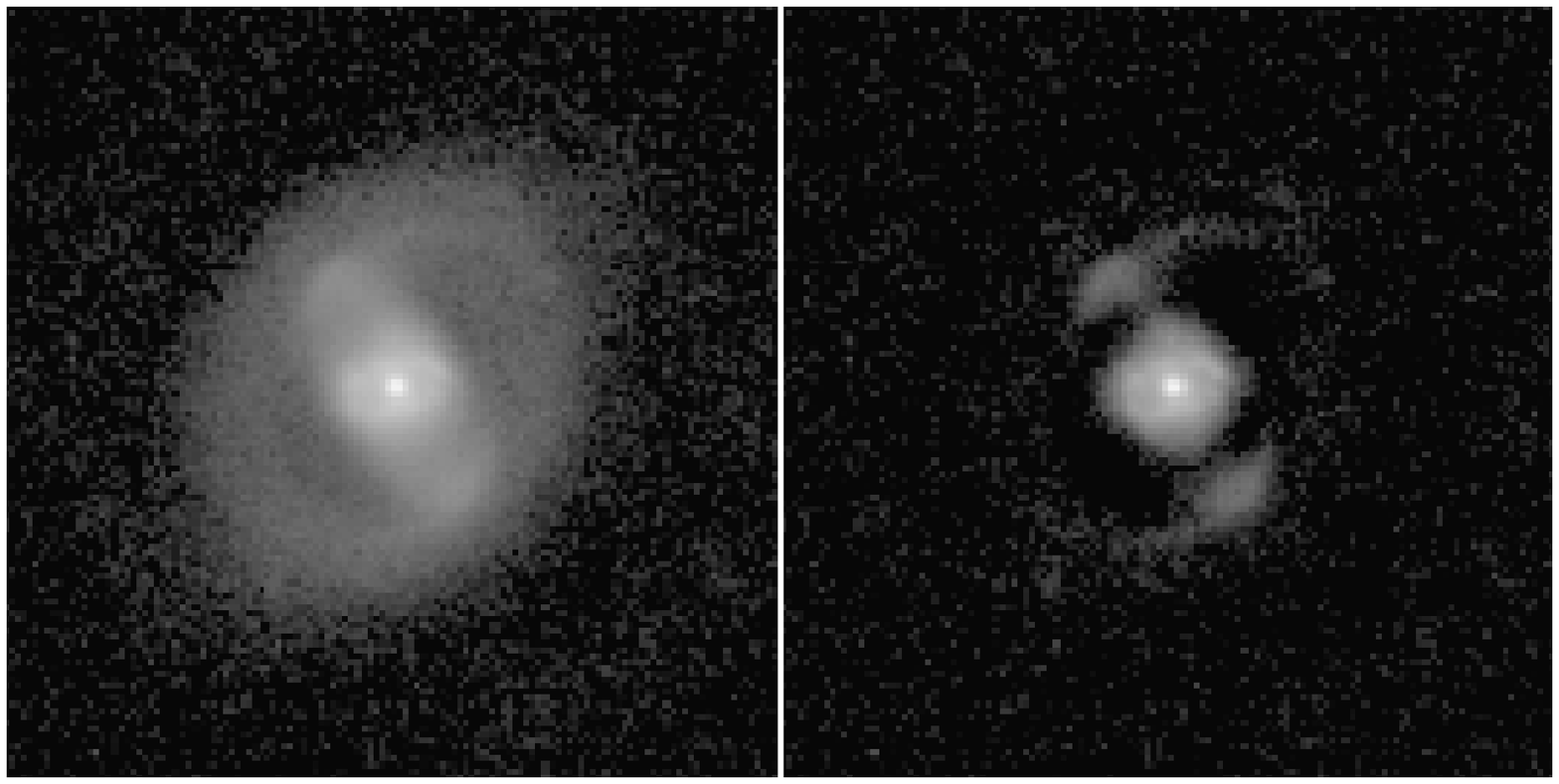}
\caption{Ansae in the early-type barred spiral ESO 565-11.
(left): $H$-band image obtained with the CTIO 1.5m telescope
(Buta, Crocker, \& Byrd 1999). Right: unsharp-masked image
of the same galaxy, showing the bright bar ansae and inner arms.
}
\label{fig:NGC2983}
\end{center}
\end{figure}

\begin{figure}
\begin{center}
\includegraphics[scale=0.5]{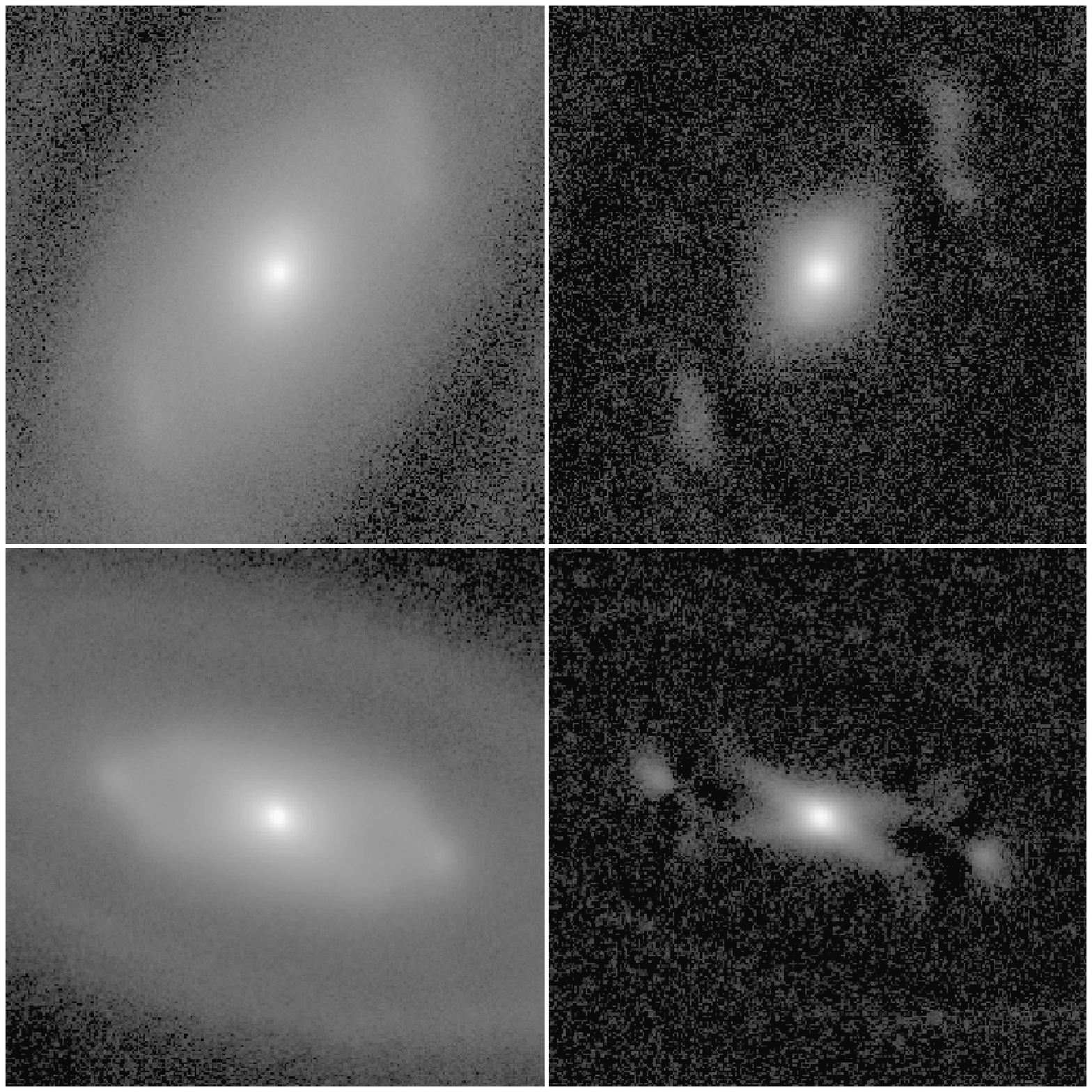}
\caption{Two unusual ansa cases. Top, left: $I$-band of NGC 7098
 (type (R$_1$R$_2$)'SAB(\underline{r}s)ab) from observations with the CTIO
 1.5-m telescope (Buta 1995). Top, right: Unsharp-masked
 version of the same image showing the mostly linear bar ansae.
 Note the slight waviness in the upper right ansa. Bottom,
 left: $I$-band image of NGC 7020 (type (R)SAB(r?)0/a) from observations 
 with the CTIO 1.5-m telescope (Buta 1995). Note the strong inner
 hexagonal zone and the bright spots on the major axis.
 Bottom, right: Unsharp-masked
 version of the same image showing two conspicuous ansae
 and a complex X-pattern in the hexagonal zone, suggesting
 a relation to bars.
}
\label{fig:NGC7020}
\end{center}
\end{figure}

Ansae can be described as a pair of density
enhancements at the ends of the bar (Fig.~\ref{fig:NGC4643}, 
Fig.~\ref{fig:NGC2983}, Fig.~\ref{fig:NGC7020}), sometimes also referred
to as ``condensations'' in older work (e.g., Danby 1965) or symmetric density knots. A more scientific definition could be, a local maximun before the end of the bar along its main axis absent on its minor axis. Ansae have been known in the literature for decades, and they appear to be rather frequent in early-type disk galaxies (e.g., NGC~4262, NGC~2859 and
NGC~2950; Sandage 1961; Athanasoula 1984). No statistical information on the prevalence of ansae has ever been published, a situation we aim to remedy with
this short paper.  Ansae were discussed by Danby (1965) in the context
of an ``outflow'' from the bar into the associated spiral arms. In
$N$-body numerical simulations, ansae are seen on both ends of a bar
as characteristic density enhancements in face-on or edge-on disks
(Martinez-Valpuesta, Shlosman \& Heller 2006). Athanassoula (2001)
related the appearance of ansae to initial conditions in the models
(e.g., the halo-to-disk mass ratio). In general, ansae in simulated
galaxies appear after some Gyrs of evolution. The dynamical
significance of ansae remains unclear. Are they regions of trapped
disk or bar particles? Do they reflect any underlying dynamics? Do
they appear in a particular stage of the bar evolution?  We will try
to answer these questions in a future paper based on numerical
simulations (I. Martinez-Valpuesta et al. 2007, in preparation,
hereafter Paper II), but first we need to stablish how frequent ansae
are and in which host galaxies they occur.

In this paper, we derive some statistics of barred galaxies with 
ansae. We present the statistics in Section~2, discuss observed
characteristics of ansae and their host galaxies in Section~3, and
discuss our results in the context of bar dynamics and evolution in
Section~4, before summarising our conclusions in Section~5. Ansae will
be considered from a dynamical point of view by means of simulations
in Paper II.

\section{Observational data and Statistics}

Our primary data source is The de Vaucouleurs Atlas of Galaxies (Buta,
Corwin \& Odewahn 2007; hereafter called the Atlas), from which we
select all the barred galaxies, as classified in the Atlas. We assume
that the sample of galaxies collected in the Atlas is random in the
sense that the presence or absence of ansae did not influence the
selection process of galaxies in the catalogue. Based on the detailed
description for each galaxy as given in the Atlas, we then separated 
all galaxies with and without ansae, and collected the statistics as a
function of morphological type. The results of this exercise are shown
in Table~\ref{tab:2007PAPERv2}. 

As a consistency check, we also use the detailed study of 26
early-type barred galaxies by Buta et al. (2006), who study the bar
strengths, bulges, disks, and bar parameters in a statistically
well-defined sample of S0-Sa galaxies. Most important for our present
paper, the authors also list whether a galaxy has ansae, as based on 
deep NIR images.

We estimate the uncertainties in all our derived fractions by using
Poisson statistics(see Laine et al. 2002).

% where the error is given by $\sigma =
%[f(1-f/N)]^{1/2}$, with $f$ the quantity that is measured and $N$ the
%size of the sample from which the quantity is derived (see Laine et
%al. 2002).%
%

In Table~\ref{tab:2007PAPERv2} we show the results of the analysis of
the sample of barred galaxies described in the Atlas. This sample
consists of $267$ barred galaxies, of which just $36$ are of ansa type
($14\%\pm 2\%$). There is, however, a very strong dependence of the
ansa fraction on morphological type, and $32$ of $36$ ansae galaxies
occur in the $120$ galaxies of type Sab or earlier ($26\%\pm 4\%$). Even
within this group of early-type galaxies we see that the highest
fraction of ansae occurs in S0 galaxies, with $14$ out of $39$
($36\%\pm8\%$). And among the lenticular galaxies, ansae are especially
prevalent among those classified as 'SB', or strongly barred, with
$42\%\pm 10\%$ of the SB0's having ansae. Only five of the $35$ Sab
galaxies ($14\% \pm 6\%$), three of the $39$ Sb, one of the $31$ Sbc
galaxies, and none of the $77$ barred galaxies in the Atlas with later
morphological types have ansae. 

In Table~\ref{tab:2006PAPER}, we show the statistics for the much
smaller sample from Buta et al.(2006), which we use as a consistency
check of the presumably unbiased sample selection in the Atlas. Buta
et al. (2006) find that ten of the $26$ barred S0-Sa galaxies in their
sample have ansae ($38\%\pm 10\%$). When considering the ansa
fraction in SB and in SAB bars, we see that $43\%\pm 13\%$ (six out of
14) of the bars in the SB0 galaxies and $20\%\pm 18\%$ (one out of five) of
those in the SAB0's have ansae.  Considering galaxies of types SB
and SAB combined, we find that $37\%\pm 11\%$ of the barred S0 and
$33\%\pm 19\%$ of the barred Sa galaxies have ansae. Within the large
uncertainties, these results are entirely consistent with those
derived, above, for the Atlas galaxies, thus confirming the validity
of the use of the Atlas for our purpose.

\begin{figure}
\begin{center}
\includegraphics[scale=0.3,angle=-90]{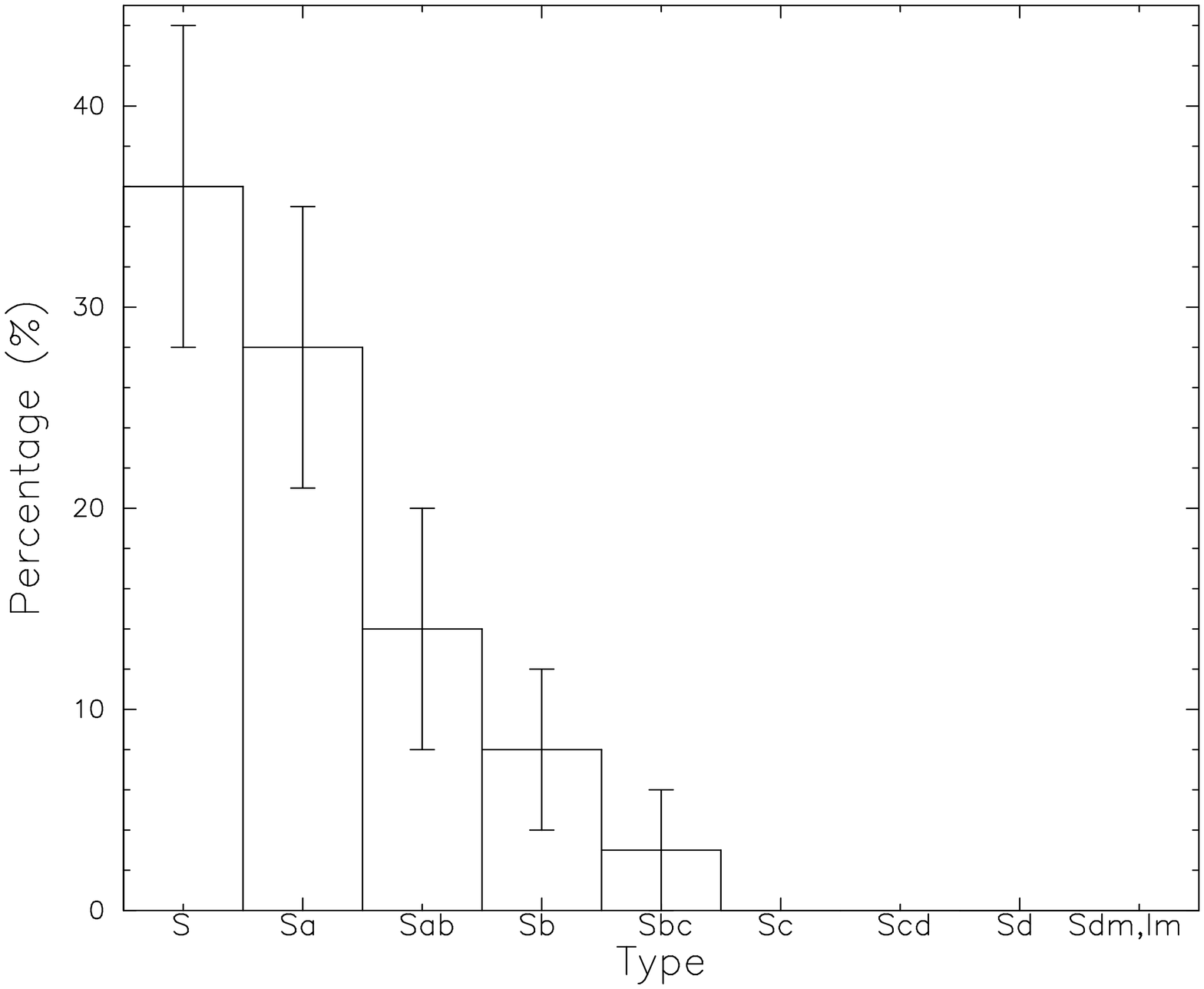}
\vspace{0.5cm}
\caption{Histogram of percentages of ansae as a function of
  morphological  type, as based on the results from the
  Atlas. Uncertainties are Poisson errors.}
\label{fig:histogram}
\end{center}
\end{figure}

The histogram presented in Fig.~\ref{fig:histogram} shows the main
conclusion we reach on the basis of the analysis described above,
namely that ansae are very common among barred lenticulars, but that
their fraction decreases very rapidly with morphological type. Ansae
hardly occur at all in galaxies of type Sb or later. We have 
carefully reviewed the images of the Atlas and still no late-types 
galaxies present ansae. We can conclude that the missing fraction of ansae in late-types is real.

\begin{table}
\begin{center}
{\small
\caption[]{Statistics for the sample of barred galaxies from the
  Atlas.}
\label{tab:2007PAPERv2}
\vspace{1.0cm}
\begin{tabular}{ l c c c c}
%\begin{tabular}{ l l r@{.}l r@{.}l }
\hline 
 Type & Total & Ansae & No ansae & Ansa fraction \\
%\multicolumn{2}{c}{$\tau$} & \multicolumn{2}{c}{$Z$} \\
\hline
\hline
 SB0 & $26$ & $11$ & $15$ & $42\%\pm 10\%$ \\
 SAB0 & $13$ & $3$ & $10$ & $23\%\pm 12\%$ \\
 SBa & $19$ & $5$ & $14$ & $26\%\pm 10\%$ \\
 SABa & $27$ & $8$ & $19$ & $30\%\pm 9\%$ \\
 SBab & $21$ & $3$ & $18$ & $14\%\pm 8\%$ \\
 SABab & $14$ & $2$ & $12$ & $14\%\pm 9\%$ \\
 SBb & $21$ & $2$ & $19$ & $10\%\pm 6\%$ \\
 SABb & $18$ & $1$ & $17$ & $6\%\pm 6\%$ \\
 SBbc & $12$ & $0$ & $12$ & $0\%$ \\
 SABbc & $19$ & $1$ & $18$ & $5\%\pm 5\%$ \\
 SBc & $6$ & $0$ & $6$ & $0\%$ \\
 SABc & $15$ & $0$ & $15$ & $0\%$ \\
 SBcd & $11$ & $0$ & $11$ & $0\%$ \\
 SABcd & $9$ & $0$ & $9$ & $0\%$ \\
 SBd & $8$ & $0$ & $8$ & $0\%$ \\
 SABd & $6$ & $0$ & $6$ & $0\%$ \\
 SBdm, & & & & \\
 IBm & \raisebox{0.6ex}{$15$} &  \raisebox{0.6ex}{$0$} & \raisebox{0.6ex}{$15$} & \raisebox{0.6ex}{$0\%$} \\
 SABdm, & & & & \\
 Im & \raisebox{0.6ex}{$7$} & \raisebox{0.6ex}{$0$} & \raisebox{0.6ex}{$7$} & \raisebox{0.6ex}{$0\%$} \\
\hline
\hline
 S0 & $39$ & $14$ & $25$ & $36\%\pm 8\%$ \\
 Sa & $46$ & $13$ & $33$ & $28\%\pm 7\%$ \\
 Sab & $35$ & $5$ & $30$ & $14\%\pm 6\%$ \\
 Sb & $39$ & $3$ & $36$ & $8\%\pm 4\%$ \\
 Sbc & $31$ & $1$ & $30$ & $3\%\pm 3\%$ \\
 Sc & $21$ & $0$ & $21$ & $0\%$ \\
 Scd & $20$ & $0$ & $20$ & $0\%$ \\
 Sd & $14$ & $0$ & $14$ & $0\%$ \\
 Sdm, %& & & & \\
Im & $22$ & $0$ & $22$ & $0\%$ \\
%& \raisebox{0.6ex}{$0\%$} \\
% Im & \raisebox{0.6ex}{$22$} &  \raisebox{0.6ex}{$0$} & \raisebox{0.6ex}{$22$}
%& \raisebox{0.6ex}{$0\%$} \\
\hline
\hline
 Total & $267$ & $36$ & $231$ & $14\%\pm 2\%$ \\
\end{tabular}
}
\end{center}
\end{table}

\begin{table}
\begin{center}
\caption[]{Statistics for the sample of early-type barred galaxies
  from Buta et al. (2006)}
\label{tab:2006PAPER}
\vspace{1.0cm}
\begin{tabular}{ l c c c c}
%\begin{tabular}{ l l r@{.}l r@{.}l }
\hline 
 Type & Total & Ansae & No Ansae &  Ansa fraction\\
%\multicolumn{2}{c}{$\tau$} & \multicolumn{2}{c}{$Z$} \\
\hline
\hline
 SB0 & $14$ & $6$ & $8$ & $43\%\pm 13\%$ \\
 SAB0 & $5$ & $1$ & $4$ & $20\%\pm 18\%$ \\
 SBa & $2$ & $0$ & $2$ & $0\%$ \\
 SABa & $4$ & $2$ & $2$ & $50\%\pm 25\%$ \\
 SBab & $1$ & $1$ & $0$ & $100\%$ \\
 SABab & $0$ & $0$ & $0$ & $0\%$ \\
\hline
\hline
 S0 & $19$ & $7$ & $12$ & $37\%\pm 11\%$ \\
 Sa & $6$ & $2$ & $4$ & $33\%\pm 19\%$ \\
\hline
\hline
 Total & $26$ & $10$ & $16$ & $38\%\pm 10\%$ \\
\end{tabular}
\end{center}
\end{table}

\section{Characteristics of bars with ansae}

Having established that ansae preferentially occur in barred
lenticular and Sa galaxies and, in the case of the former where the
number statistics are best, more in bars of SB type than in those of
SAB type, in this Section we will explore relations with the strength
of the bar and with the presence of rings in the host galaxies, 
and describe further characteristics of the ansa regions.

\subsection{Bar strength}

The strength of bars can be measured by deriving the gravitational bar
torque, indicated by the parameter $Q_{\rm b}$ (Sanders \& Tubbs 1980;
Combes \& Sanders 1981; Buta \& Block 2001; Laurikainen et al. 2004). 
This parameter describes the maximum gravitational bar torque per
unit mass per unit square of the circular speed as inferred from 
near-IR images under the assumption of a constant mass-to-light ratio.
In intermediate to late-type
spirals it is essential to correct the {\it total} non axisymmetric
strength $Q_g$ for spiral arm torques in order to derive $Q_{\rm b}$,
while for early-type galaxies
one can set $Q_{\rm b}$ $\approx$ $Q_g$ (Buta et al. 2005).
$Q_{\rm b}$ can be affected by the strong axisymmetric
background in early-type barred galaxies, and may be significantly
diluted by a bulge or inner disk component.

Buta et al. (2006) list $Q_{\rm b}$ for all the 26 galaxies in their
sample, as derived from 2.15$\mu$m $K_s$-band images (no sufficient $Q_{\rm b}$
measurements are available in the literature to perform a similar
analysis for all the galaxies in the Atlas sample). The median value of
$Q_{\rm b}$ for the early-type barred galaxies with ansae is $0.207$
($\sigma=0.07$), compared to a median $Q_{\rm b}$ of $0.127$
($\sigma=0.1$) for those without ansae. We have performed the Kolmogorov-Smirnov test, and the difference between ansaed barred galaxies and the non ansaed barred galaxies is statistically significant with $p=0.05$. The ansa bars are thus
stronger than non-ansa bars, in the median. 

\subsection{Rings}

We have also studied the ansae in relation to the occurrence of inner
and outer rings in their host galaxies. Based on the morphological
classification of each galaxy, we collected the statistics for the
early-type galaxies (SB0,SB0/a,SAB0,SAB0/a, where we find the ansae) in
both of our samples. We find similar fractions of ringed galaxies
among those that host ansae and those that do not. This indicates that
the ansae are related purely to the bar, and presumably to its
strength (see above), and that the mechanism producing the ansae is
unrelated to the mechanism producing the inner or outer rings. So
although both ansae and rings (see the review by Buta \& Combes 1996)
are intimately related to the bars of the galaxies that host them, our
findings show that the two phenomena are not related among them. Ansae
and rings trace different consequences of bar evolution and dynamics,
which can coincide in some galaxies.

\subsection{Properties of ansae regions}

By carefully looking at the data of our galaxy sample, we are able to give an overview of the characteristics of ansae. 
In Fig.~\ref{fig:NGC7020}, we present examples of different barred galaxies showing clear density enhancements at the end of the bar, using in some cases the technique of unsharp-masking (low pass filtering) to make
the features more obvious. The standard examples of ansae are represented by Figs.~\ref{fig:NGC2983}. 
Figure~\ref{fig:NGC4643} shows a {\it non-ansae} (regular) bar (NGC 4643) for comparison with a typical ansae
bar (NGC 936) to highlight that a real distinction is present. Figure~\ref{fig:NGC7020} shows two special
cases that are not necessarily unique but which are clearly unusual. In addition to two round ansae,
NGC 7020 has an X-shaped feature in its unsharp-masked image, while NGC 7098 shows two elongated ansae
with a clear wavy appearance.

\begin{figure}
\begin{center}
\includegraphics[scale=0.5]{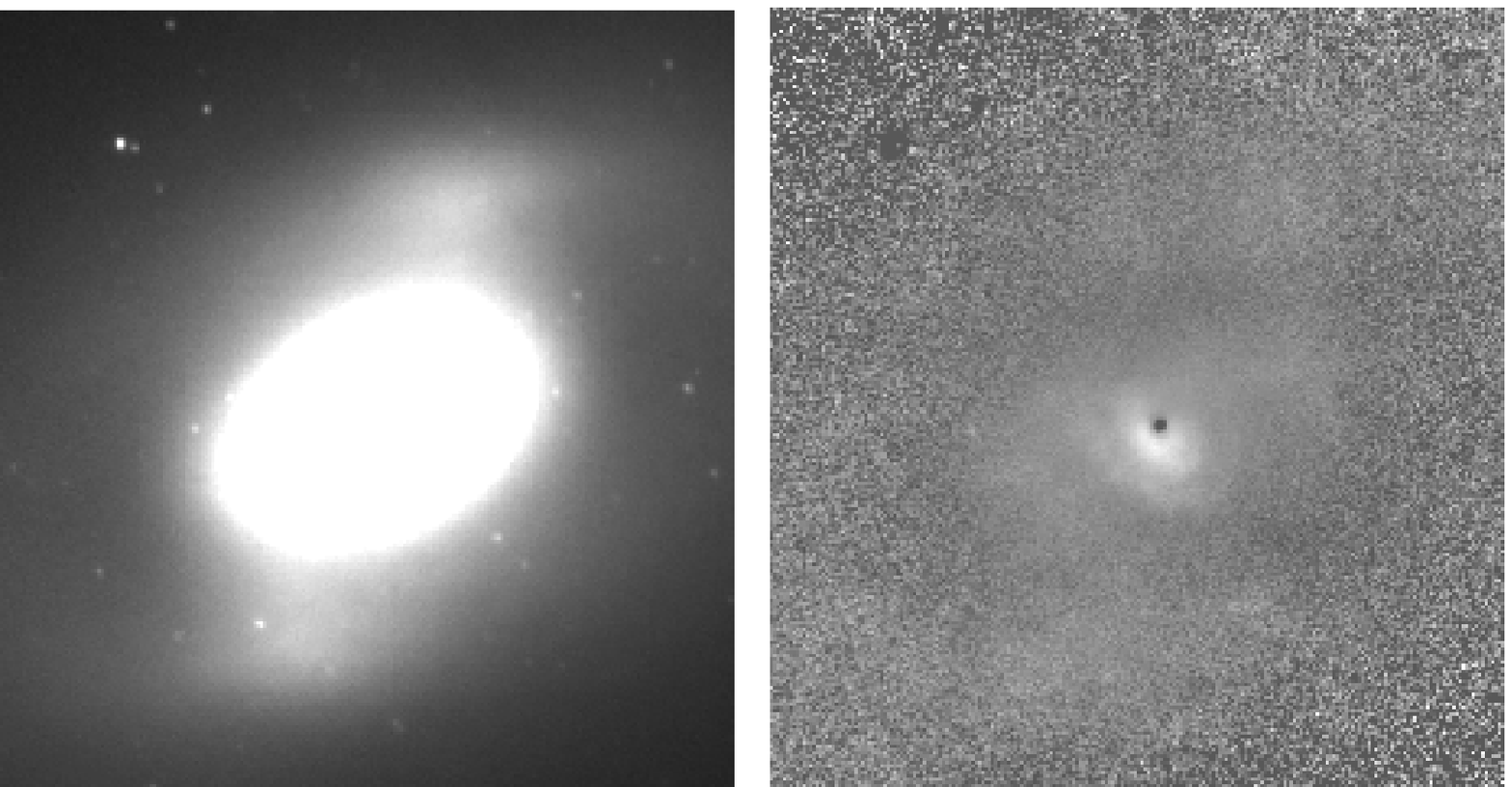}
\caption{{\it Left}: $R$-band and, {\it right}, $B-R$ color index images of
NGC~2787 (type SB(r)0$^+$), where darker shades indicate redder colors in the latter. The
images were obtained in service time on the night of 2007 February ninth
with the ALFOSC camera on the 2.5\,m Nordic Optical Telescope. Images are
about 80\,arcsec on the side, N is up and E to the left.
}
\label{fig:NGC2787}
\end{center}
\end{figure}

\begin{figure}
\begin{center}
\includegraphics[scale=0.5]{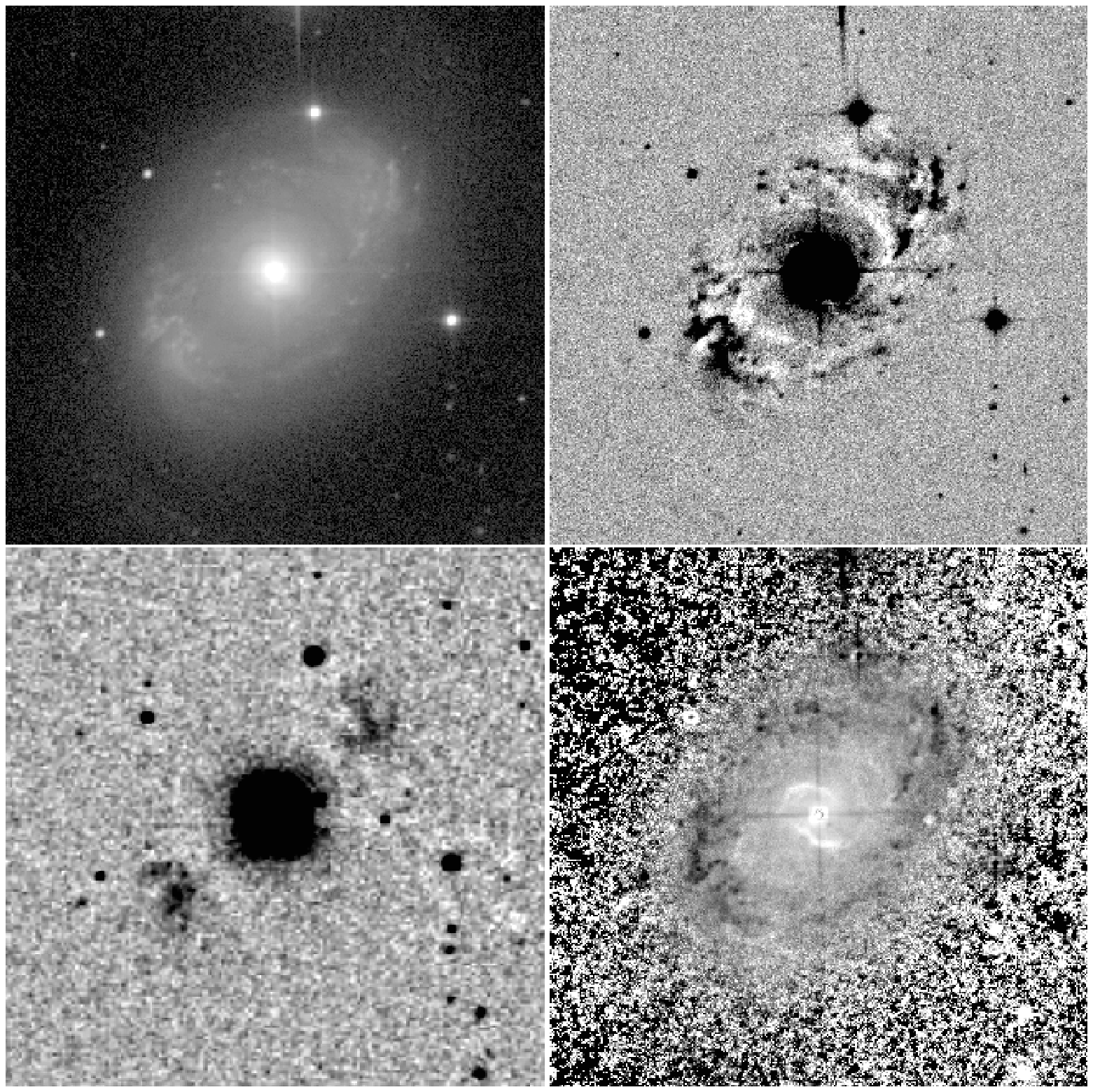}
\caption{The bar/oval region of the well-known Seyfert 1
galaxy NGC 4151 (type (R$_2$)'SAB(s)ab) based on images from Eskridge et al. (2002).
Upper left: $B$-band (units
mag arcsec$^{-2}$); upper right: unsharp-masked $B$-band
image; lower left: unsharp-masked $H$-band image; lower
right: B-H colour index map coded such that blue regions are
dark and red regions are light. The images show that the
ansae in NGC 4151 are partly defined by star-forming
regions, and that the star formation may be part of an
intrinsically elliptical inner ring lying at the boundary
of the inner oval.
}
\label{fig:NGC4151}
\end{center}
\end{figure}

An interesting characteristic of ansae is that the majority of our galaxies do not present signs of star formation in these regions (Buta et al 2007). 
In colour index maps such as those shown in the Atlas, ansae typically show no colour enhancement, indicating
they must be largely stellar dynamical phenomena (e.g., NGC 2787, Fig.~\ref{fig:NGC2787}). The only exception we know of is the well-known
Seyfert 1 galaxy NGC 4151 (Fig.~\ref{fig:NGC4151}), where the ansae appear in blue light to be collections
of star-forming regions. The $H$-band image of NGC 4151 also shows a significant stellar component
to its ansae. Both the $B$- and $H$-band unsharp-masked images show the associated groups of star-forming
regions, but the ansae themselves do not stand out strongly in the $B-H$ colour index map. Instead,
this map suggests that the star formation is related to an elliptical inner ring where the HII regions
concentrate around the bar ends (Crocker, Baugus, and Buta 1996). 

\section{Discussion}

Danby (1965) published the first explanation of ansae, namely as an
outflow from the bar into the spirals. Since
Danby's paper, hardly any work has been published on this kind of bar
feature. In recent years, simulations have clearly shown ansae in
modeled barred galaxies (Athanassoula 2001; Athanassoula \ Misiriotis 2002; 
Martinez-Valpuesta \ea 2006; Martinez-Valpuesta 2006, especially her Fig.~5.3). 

Unlike in any kind of observations, in numerical simulations the
position and velocity of a single particle can be traced at different
times. This technique allows the modeller to visualise the stellar
orbit of a particle which is currently located in the ansae region,
leading to an insight into where it comes from and where it will go.
Martinez-Valpuesta (2006) found that particles get trapped in the
region of the ansae, but, later on, continue to form part of the bar.
A more detailed study of these orbits and the implications will be
presented in Paper~II. There, we will give a statistical analysis of
the ansa region and quantify the amount of particles (and hence the
stellar mass) that populate the ansa region.

Above, we have seen that ansae appear mainly in lenticulars,
exclusively in early-type galaxies and not at all in the later types.
This dependence on morphological type will be explored in more detail
in Paper~II. The observational results, as outlined above, also show
that the bar strength, as measured through the gravitational bar
torque, is higher in the median for ansa bars than for non-ansa
bars.

The fact that ansae are preferentially found in strongly barred
early-type galaxies is reproduced in the simulations (Paper~II), and
the inference drawn there is that ansae are present when the bar is
both strong and growing. In paper II, we will attempt to present a
theoretical and numerical explanation of the ansae. The use of
hundreds of simulation runs allows us, firstly, to produce statistics
on the bars and the ranges of model parameters which lead to the
occurrence of ansae. Secondly, we will perform a study of the dynamics of
the stellar orbits in the ansa region, yielding statistical
information on the amount of particles (tracing the mass) that is
trapped in the ansae at any given moment, or that passes through the
ansae before becoming part of the bar.

Debattista \& Sellwood (2000) and later, Athanassoula \& Misiriotis (2002) 
showed that bars formed in galaxies with more concentrated halos are 
stronger than those that formed in less 
concentrated ones. This was explained by Athanassoula (2003), 
who showed that a more concentrated halo has more mass at the relative 
resonances (particularly CR) and thus leads to more angular momentum 
exchange within the galaxy, and thus to a stronger bar.
 If we extrapolate this, we
could state that the galaxies presenting ansae are those that are in
the process of secular evolution due to the dynamics of the bar. It
would be worthwhile to investigate this proposition further
observationally, for instance by analysing the relative contributions
of disk and bulge components in the inner parts of radial surface
brightness profiles. This is beyond the scope of the present paper.

\section{Conclusions}

We have assessed the frequency of occurrence of ansae, or the
``handles'' of the bar, in barred galaxies of various morphological
types. Although such ansae have been described in the literature for
decades, there is as yet no information at all on how common they
are. We base our analysis on classifications given by Buta et
al. (2006) for a sample of 26 early-type barred galaxies, and in the
Atlas for a much larger sample of 267 barred galaxies, but of all
types. 

Our main conclusion is that ansae never occur in late-type galaxies
(types Sc or later) and only very rarely in intermediate (Sab-Sbc) types,
but that they are very much concentrated in early-type galaxies. The
highest fraction of ansae is found in strongly barred lenticulars,
with $\sim40\%$ of SB0 galaxies showing ansae. The overall fraction
among S0 and Sa galaxies is around a third.

We also find that the median bar strength, as based on literature
measurements of the gravitational bar torque, is significantly higher
among barred galaxies with ansae than among those without. The
fraction of inner and outer rings among ansa and non-ansa bars is
not different, so although both ansae and rings are related to bars
they are not related through the same mechanism.

We conclude that ansae are common in early-type, but absent from
late-type barred galaxies, and tend to occur in galaxies with strong bars. In
our subsequent Paper~II, we will explore the origin and properties of
ansae, and relate them to the evolution of the host galaxy and bar
through numerical simulations.

\section*{Acknowledgements}

We thank Drs. H. Salo and E. Laurikainen for the use of their
multicomponent decomposition code. We thank E. Athanassoula for comments 
on an earlier version of this manuscript. This work has been partially 
supported by the Peter Gruber Foundation Fellowship. RB acknowledges 
the support of NSF grant AST050-7140. Funding for the creation and
distribution of the SDSS Archive has been provided by the Alfred P. Sloan
Foundation, the Participating Institutions, NASA, NSF, the U.S. Department
of Energy, the Japanese Monbukagakusho, and Max Planck Society.

\label{lastpage}

\end{document}